# Gate-Controlled Ionization and Screening of Cobalt Adatoms on a Graphene Surface

Victor W. Brar[1,2,*], Regis Decker[1,2,*], Hans-Michael Solowan[1,*], Yang Wang[1], Lorenzo Maserati[1], Kevin T. Chan[1,2], Hoonkyung Lee[1,2], Caglar O. Girit[1,2], Alex Zettl[1,2], Steven G. Louie[1,2], Marvin L. Cohen[1,2], and Michael F. Crommie[1,2]

[1]Department of Physics, University of California at Berkeley, Berkeley, CA 94720

[2]Materials Sciences Division, Lawrence Berkeley Laboratory, Berkeley, CA 94720

*These authors contributed equally to this work

**The accessibility of graphene's surface provides a unique opportunity to modify its electronic properties via the deposition of chemical impurities. Adsorbates on graphene can be used to tune electronic scattering in graphene [1,2], alter electron-phonon interactions [3,4], shift the chemical potential [1-3,5,6], change the effective dielectric constant [7], and – in cases such as 'graphane' [8] – form whole new 2D materials. Understanding the microscopic physics of such behavior requires local-probe exploration of the subnanometer-scale electronic and structural properties of impurities on graphene. Here we describe scanning tunneling spectroscopy (STS) measurements performed on individual cobalt (Co) atoms deposited onto backgated graphene devices. We find that Co adatoms on graphene can be ionized by either the application of a global backgate voltage or by the application of a local electric field from a scanning tunneling microscope (STM) tip. Large screening clouds are observed to form around Co adatoms ionized in this way, and we observe that some intrinsic graphene defects display a similar behavior. Our results provide new insight into charged impurity scattering in graphene, as well as the possibility of using graphene devices as chemical sensors.**

Graphene impurities provide both a source of mobility-limiting disorder as well as a means to alter graphene electronic structure in a desirable way. Thus far, the effects of impurities on graphene have been primarily studied using spatially averaged techniques such as transport measurements [1,2,5], Raman spectroscopy [9], and Angle-Resolved Photoemission (ARPES) [3,6]. Models have been developed to account for graphene impurity behavior based on these non-local experiments [1,10,11], but systematic local-probe measurements of impurities on gated graphene samples have not yet been performed. In other material systems, however, the STM has proven to be a useful tool to probe charged impurities on insulating and semiconductor surfaces, providing Ångstrom-scale measurement of ionized impurities and associated screening potentials [12,13]. Graphene differs from these previously studied systems in that it is a semi-metal surface with a charge density that can be electrostatically tuned via application of a backgate voltage. Our measurements show that, despite the metallicity of graphene, impurities on its surface can be charged and discharged through the application of a backgate voltage, thus affecting the overall graphene behavior.

Our experiments were performed using an Omicron LT-STM operating in ultra high vacuum (UHV) ($<10^{-11}$ torr) and at 4.2 K. STM tips were calibrated spectroscopically against the Au(111) Shockley surface state before all measurements. STM differential conductance (dI/dV) was measured by lock-in detection of the ac tunnel current modulated by a 1−5 mV (rms), 350−500 Hz signal added to the tunneling bias $V_b$ ($V_b$ is the voltage of the sample with respect to the tip). The dI/dV($V_b$) and I($V_b$) measurements were performed under open feedback conditions with the tip in a fixed position. Single graphene sheets were prepared using either mechanical exfoliation [14] or by chemical vapor deposition (CVD) [15]. In both cases the graphene sheets were placed on a 285 nm thick $SiO_2$ layer grown atop a heavily doped Si crystal which was used as a backgate electrode [14]. Electrical contact was made to the graphene by direct deposition of Ti (10 nm thick) / Au (30 nm thick) electrodes through a stencil mask. Samples were annealed at temperatures between 200 and 400 °C for several hours in UHV to clean them before cooling. Co adatoms were deposited via e-beam evaporation onto pristine, cold (<10 K) graphene surfaces displaying large areas (e.g., 20 × 20nm$^2$) with no observable defects. Figure 1 shows a representative $10 \times 10$ nm$^2$ scan of a graphene surface following Co adatom deposition. Co adatoms appear as ~4 Å high dome-like protrusions on the graphene surface (two cobalt adatoms can be seen in this image). In order to understand the possible role of defects in our measurements, defects were introduced into some graphene surfaces through high temperature annealing (400−500 C) (a topographic image of a typical defect can be seen in Fig. 3f, inset). We observe that the presence of nearby defects did not change the behavior of subsequently deposited Co adatoms. Furthermore, atomic manipulation was utilized to insure that adatoms studied were not bound to defect sites.

We performed dI/dV measurements with the STM tip held over individual Co adatoms in order to measure their energy-dependent local density of states (LDOS). Figure 2a shows typical dI/dV spectra obtained from a single Co adatom for different applied back-gate voltages, $V_G$ (changing $V_G$ changes the graphene charge carrier density). Three different types of features are observed in these spectra. The first is a dip observed at the Fermi level ($E_F$) having a full-width at half maximum (FWHM) of ~10 meV (see inset for a higher resolution spectrum). The energy and width of this feature does not vary significantly with $V_G$. The second type of feature is a series of resonant peaks in the dI/dV signal marked 'A', 'B', 'C' and 'D' in Fig. 2a. These peaks have a typical FWHM of ~25 ± 5 meV. Their behavior differs from the dip at $E_F$ in that their energy locations move in the same direction (and by similar amounts) as the Dirac point voltage ($V_D$) as $V_D$ is varied by application of $V_G$ ($V_D$ marks the location of a dip seen in the graphene LDOS which is associated with the Dirac point energy [16]). The third type of feature corresponds to the peak in dI/dV signal marked 'S' in Fig. 2a. The energy of this feature changes as $V_G$ is varied, but it shifts energetically in the opposite direction compared to $V_D$, and it disperses significantly faster. Figure 2b summarizes how these spectral features shift in energy as a function of $V_G$. Also plotted in Fig. 2b is the gate-dependent variation in Dirac point voltage ($V_D$) measured on the graphene surface at least 10 nm away from any Co atom.

In order to better understand how the presence of a Co adatom affects the behavior of graphene, dI/dV measurements were also performed on the bare graphene surface adjacent to a Co atom. Figure 3a shows dI/dV($V_b$) and I($V_b$) signals measured on the graphene surface at a lateral distance of ~2.5 nm from the center of a Co adatom. The dI/dV spectrum shows a ~126 meV wide gap-like feature at $E_F$ as expected for a clean graphene surface (this is known to arise from inelastic tunneling [16]), but an additional sharp dip is also observed ~220 meV above $E_F$. We measured how this new dI/dV feature varies spatially through the use of dI/dV mapping. Figure 3b−d shows dI/dV maps acquired in the vicinity of a single Co atom with different $V_G$ and $V_b$. The most striking aspect of these images is a narrow ring, arising from the spectral dip feature above $E_F$, that appears centered around the Co adatom. The diameter of this ring is strongly dependent on $V_b$ and $V_G$. For example, Figure 3c,d shows that the ring diameter increases as $V_G$ is decreased, while Figure 3b,d shows that the diameter also increases as $V_b$ is increased.

Some dI/dV maps were also acquired over the bare graphene surface (i.e., in the absence of cobalt adatoms) after annealing to higher temperatures (400–500 °C) as seen in Fig. 3e. Ring-like features similar to those found around Co adatoms can be seen. The inset of Fig. 3f shows an STM topograph of the center of one of these rings, revealing a defect in the graphene. A dI/dV spectrum taken at the site of such a defect (Fig. 4f) shows two clear spectroscopic features, R' and S'. R' is observed to disperse in the same direction as $V_D$ when $V_G$ is varied, while S' disperses in the opposite direction. Similar behavior was seen for numerous defects on the graphene surface in the absence of cobalt adatoms.

The spectroscopic features we observed for cobalt adatoms and intrinsic graphene defects can be understood within a general impurity physics framework, which we now describe. We start with the dip feature at $E_F$ observed for cobalt adatoms. This is not a conventional band structure feature because it remains pinned to $E_F$ even as $E_F$ is swept through the graphene bandstructure by backgating. One possible origin of this feature is the Kondo effect, which arises from spin-screening of a local moment by itinerant substrate electrons [17]. The Kondo effect results in a resonance at $E_F$ that often manifests as a dip in LDOS at the site of a magnetic impurity [18]. Cobalt is expected to be magnetic on graphene [19], and in this interpretation the Kondo temperature extracted from the dip feature is ~60 K [20]. However, a Kondo resonance is generally expected to have a strongly gate-dependent width [21], which is not observed here. This opens the possibility that the dip at $E_F$ arises from another origin, such as vibrational inelastic electron tunneling (IET) [22]. IET causes an increase in dI/dV conductance at vibrational energy thresholds for both positive and negative biases, and so can produce dip-like features in dI/dV spectra with a half-width equal to the vibrational mode energy[22]. In order to test this possibility, we performed *ab initio* calculations of a Co adatom on graphene using a $CoC_{32}$ unit cell with the Co adatom sitting in a hollow site of the carbon lattice (calculation details are similar to Ref. [23]). Our simulation indicates that this system has in-plane vibrational modes of ~11 and 18 meV, as well as out-of-plane modes of ~15, 39 and 55 meV. The lowest energy mode is similar in magnitude to the 5 meV half-width of the dip we observe for cobalt adatoms, suggesting an IET origin for this experimental feature.

We now turn to the cobalt-induced resonant peak features marked A, B, C and D in Figs. 2a,b. We identify these features with the impurity DOS of the combined cobalt/graphene electronic structure because they shift energetically in the same direction as the Dirac point voltage ($V_D$) as $E_F$ is swept through the graphene bandstructure. Figure 4a(i) and 4b(i) show a sketch of how an impurity-induced DOS feature can be expected to shift as a result of applied gate bias. A likely explanation for these resonances (due to their narrow energy width and spatial localization) is that they arise from a hybridization of discrete cobalt atomic levels with graphene continuum states. Such resonances have been predicted for transition-metal atoms on graphene, [19,23] but our observed energy level multiplicity and spacings do not match those predictions. Another possible explanation for the observed multiplicity is hybrid electronic-vibrational (i.e., vibronic) [24] impurity states which are expected to have an energy spacing equal to the energy of a vibrational mode (our experimentally observed spacing is on the order of the energies we calculated for the out-of-plane Co/graphene vibrational modes). Lastly we point out that these states might be related to predicted fluctuations in LDOS due to screening of a 'super-critical' Coulomb impurity on graphene (i.e. quasi-Rydberg states predicted in an 'atomic collapse' scenario) [25].

Regardless of their origin, the Co atom DOS features can be either emptied or filled with electrons as they are moved above or below $E_F$ through application of a gate voltage. The energy position of these resonances with respect to $E_F$ determines the ionization state of a Co atom, and the Co ionization state can thus be externally controlled via application of a gate voltage. This interpretation is supported by the existence of both the S-state and the ring structure surrounding each atom. We first discuss the S-state, which moves opposite in energy compared to both $V_D$ and the resonant peaks when a gate voltage is applied (Fig. 2a,b). This behavior is the reverse of what is expected for a typical DOS feature, but makes sense in the context of cobalt adatom ionization [13]. Here the application of a tip bias (as well as any difference in tip-sample work functions) causes a local gating of the sample under the tip which is added to the more global gating caused by the backgate electrode, thus inducing the impurity DOS to rise ($V_b > 0$) or fall ($V_b < 0$) with respect to $E_F$. The Co atom becomes ionized when the tip bias is great enough to cause impurity states to cross $E_F$, thus creating a screening-induced response in the dI/dV signal (i.e., the S-peak). As suggested in the sketches of Fig. 4, this mechanism works equally well in the case when the impurity state is placed above $E_F$ by the backgate ($V_G < -35V$) and when the impurity state is held below $E_F$ ($V_G > -35V$), except that the S-peak is seen on opposite sides of $E_F$ for these two cases. Such behavior can be seen in the data of Fig. 2a,b where the resonant peaks and S-state lie on opposite sides of $E_F$ and (in the case of state 'A') even cross $E_F$ at the same $V_G$. This ionization framework also explains the behavior of the R' and S' resonances seen for graphene defects (the R' and S' state of the defect play the same rale as the A and S states, respectively, of cobalt).

Such ionization behavior can be analyzed within a simple double-gate model (i.e., graphene plus backgate plus tip-gate). Here we fix the backgate voltage ($V_G$) and calculate the local band-bending arising from the electric potential difference (as well as the work function difference) between the STM tip and the graphene substrate beneath it (i.e. we calculate the change in the local graphene electronic density due to tip-sample capacitance) [26]. If we roughly model the tip as a flat electrode having a work function $\Phi_{tip} = 4.8$ eV [27] and held 6 Å above the graphene surface, then, for $V_G = -40$ V, we calculate that an applied potential difference of $V_b = -105$ mV will cause impurity ionization for states observed at $V_b = +45$ mV or less. This correlates well with our STS measurements of the Co adatom for $V_G = -40$ V, where we observe an impurity state ('A') at $V_b = +44$ mV and an ionization peak ('S') at $V_b = -100$ mV. While this model provides a good framework for understanding our data, the accuracy is limited at present since the expected band-bending is a nonlinear function of $V_G$, $V_b$, work function difference, and tip height. Changing parameters such as $\Phi_{Tip}$ or the tip-height by ~30% could easily change the calculated threshold of ionization by a factor of 2 (see Supplementary Materials).

The gate-dependent ionization of cobalt atoms explains the ring structure seen surrounding each atom in dI/dV maps and topography (Fig. 3). As shown in the sketch of Fig. 4c, when the tip is displaced laterally away from an atom its electric field can still induce a local gating that shifts cobalt impurity states with respect to $E_F$, causing ionization [28]. This ionization leads to a change in LDOS surrounding the atom arising from a screening charge, and a resulting change in the measured dI/dV at the radius, $r_0$, where the ionization takes place. The cobalt atom is ionized when the tip is held at a lateral distance $r < r_0$ for such a bias and gate voltage (i.e., tip inside the ring), whereas for $r > r_0$ the atom is not ionized (i.e., tip outside the ring). The size of $r_0$ is dependent on $V_b$, $V_G$, tip height, and tip shape. For example, if $V_G$ is changed such that the impurity

DOS features are further from $E_F$, then the STM tip will need to be brought closer to the Co adatom to ionize it. This expected behavior corresponds to the experimental behavior shown in Fig. 3b−d. This type of behavior also explains the appearance of ring structure around graphene defects, as seen in Fig. 3e,f, indicating that they also exhibit gate-induced ionization. Tip-induced ionization rings have previously been seen in systems exhibiting an energy gap such as impurities on $C_{60}$ films [13] and dopants in semiconductor systerms.[12]

The rings we observe in graphene around cobalt atoms are indicative of a screening cloud that surrounds each charged adatom. We observe these rings to have diameters greater than 12nm, indicating that the screening clouds can be quite large and should likely affect macroscopic transport measurements (charged impurity scattering in general has already been proposed as a major limitation of graphene mobility [1,10]). The positioning of localized graphene impurity and defect states either above or below the Dirac point thus can explain asymmetries in graphene electron and hole conduction observed in recent transport measurements [1,2]. Although often viewed as a problem, such behavior could potentially be utilized to tune the sensitivity of graphene chemical sensors based on chemical adsorbate electronic structure.

In conclusion, we have performed gate-dependent measurements of individual Co adatoms on a graphene surface. We find that adatoms display a rich, gate-dependent electronic structure and can be reversibly ionized by changing backgate voltage. Ionized adatoms are seen to form large screening clouds that can be visualized in graphene. Similar ionization behavior is also observed for intrinsic graphene defects.

**Figure Captions:**

**Figure 1: 10×10 nm² STM topograph showing two cobalt adatoms resting atop a graphene sheet on a SiO$_2$ substrate** (tunneling parameters: $V_b = -0.25$ V, I = 10 pA, $V_G = 0$ V).

**Figure 2: Gate-dependent dI/dV spectra of Co adatom on graphene. a,** dI/dV spectra taken with tip directly above a cobalt adatom on graphene for different back-gate voltages ($V_G$) (initial tunneling parameters: $V_b = +0.15$ V, I = 22 pA). Carrier-density-dependent spectral features are labeled 'A', 'B', 'C', 'D' and 'S'. Inset: High resolution dI/dV spectrum of dip-like feature at Fermi level (initial tunneling parameters: $V_b=-0.08$ V, I= 8 pA, $V_G= -50$ V). **b,** Gate voltage dependence of dI/dV spectral features observed in (a). Open circles indicate Dirac point voltage ($V_D$) measured 10nm from cobalt atom.

**Figure 3: dI/dV conductance measurements of graphene surface near Co adatoms and surface defects. a,** Off-atom dI/dV($V_b$) and I($V_b$) spectra of graphene surface with tip held a lateral distance ~2.5nm from the center of a cobalt adatom (initial tunneling parameters: $V_b = +0.4$ V, I = 10 pA, $V_G = -40$V). **b-d,** dI/dV maps of a single cobalt adatom and surrounding graphene surface for varying $V_b$ and $V_G$ values. The 'X' in (b) indicates the STM tip position for spectrum shown in (a). **e,** dI/dV map of a bare graphene surface (i.e. without cobalt adatoms) that has been annealed at 500 °C in vacuum. Observed ring-like features are centered on defects in the graphene (tunneling parameters: $V_b=+0.7$ V, I=8 pA, $V_G= -4$ V). **f,** dI/dV spectrum taken with tip held over defect site found at the center of a defect-induced ring such as those seen in (e). Two prominent features are observed, labeled R' and S' (initial tunneling parameters: $V_b=-0.5$ V, I=5 pA, $V_G=-2.5$ V). Inset: STM topograph of defect site found at the center of a ring-like feature similar to those shown in (e) (tunneling parameters: $V_b=-0.3$ V, I=10 pA, $V_G=-60$ V).

**Figure 4: Schematic of tip and backgate-induced ionization of adatom on graphene.** **a(i)**, For $V_b=0$ and $V_G \sim 0$ V the adatom impurity DOS feature (i.e., resonance) is below $E_F$ and the adatom is neutral. **a(ii),** When a negative voltage is applied to the STM tip ($V_b > 0$) the graphene bands shift up relative to $E_F$ (i.e. there is a reduction in the local electron density of the graphene). A large enough shift pulls the impurity state above $E_F$, thus changing the adatom charge state. **b(i),** For $V_b=0$ and $V_G < -35$ V the Co impurity state is above $E_F$ and the Co adatom is charged (ionized). **b(ii),** When a positive bias is applied to the STM tip ($V_b < 0$) the graphene bands shift down relative to $E_F$. A large enough shift will push the impurity state below $E_F$, thus returning the cobalt atom to a neutral state. **c,** Schematic showing positional dependence of the band-bending induced by an STM tip. In this figure the Co adatom is neutral (i.e., the impurity state is below $E_F$).

**For methods:**

These measurements were reproduced on more than 50 Co adatoms measured with 10 different PtIr tips on 5 different graphene devices (3 exfoliated, 2 CVD grown). Although the majority of atoms measured in this work displayed the behavior described here, a fraction (~30%) of atom-like protrusions observed in our measurements strayed from this behavior in non-systematic ways. We attribute these to either Co dimers and trimers that formed as a result of diffusion during evaporation, or to Co adatoms that have attached to defects in the graphene surface.

**Acknowledgements:**

We thank J. Repp for useful discussions. Research supported by the Director, Office of Science, Office of Basic Energy Sciences, Materials Sciences and Engineering Division, of the U.S. Department of Energy under contract No. DE-AC02-05CH11231 (STM instrumentation development and operation), by the Office of Naval Research MURI Award No. N00014-09-1-1066 (experimental data analysis), and by National Science Foundation Grant Nos. DMR-0906539 (graphene device synthesis) and DMR-0705941 (electronic structure calculation). Computational resources have been provided by DOE at Lawrence Berkeley National Laboratory's NERSC facility and the Lawrencium computational cluster resource provided by the IT Division at the Lawrence Berkeley National Laboratory. H.–M. S. was partially supported by the German Academic Exchange Service.

at Lawrence Berkeley National Laboratory's NERSC facility and the Lawrencium computational cluster resource provided by the IT Division at the Lawrence Berkeley National Laboratory. H.–M. S. was partially supported by the German Academic Exchange Service.

**Figure 1.**

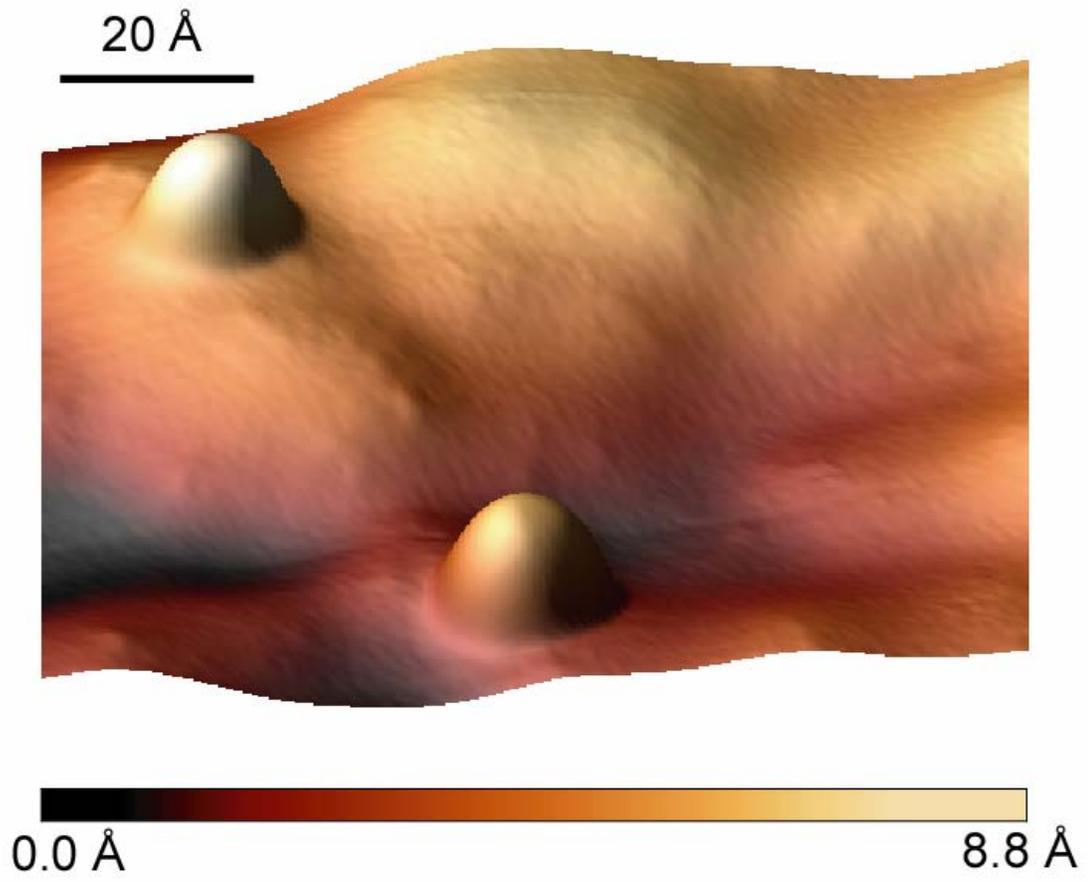

**Figure 2.**

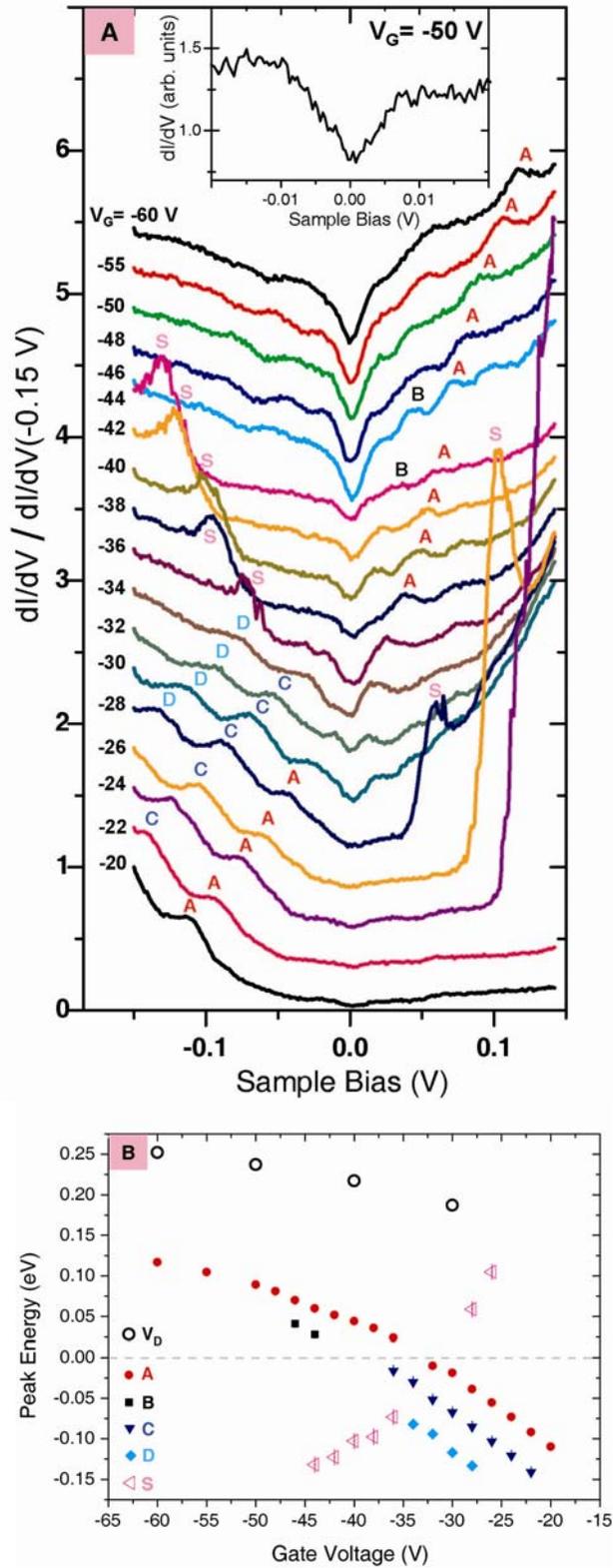

**Figure 3.**

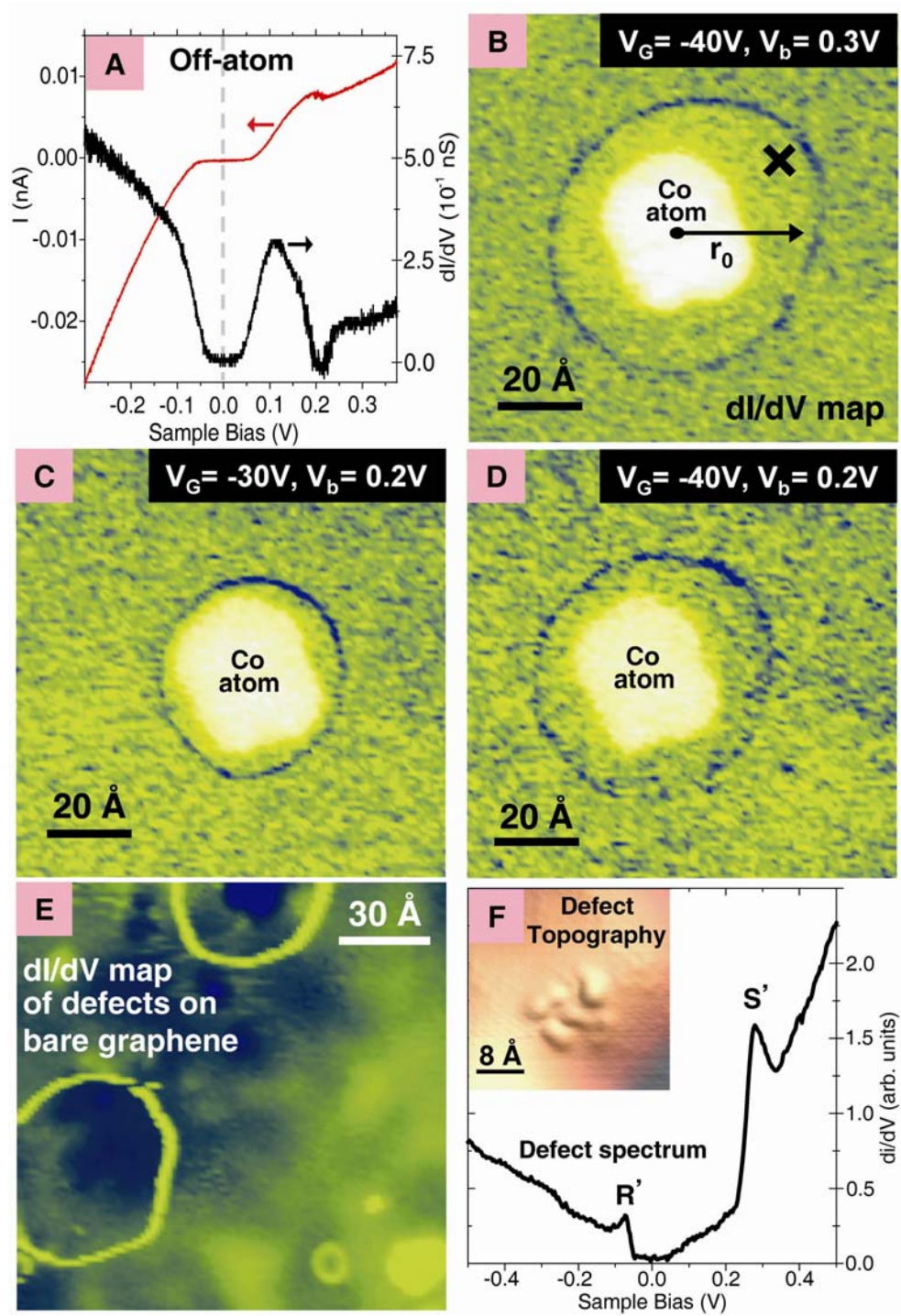

**Figure 4.**

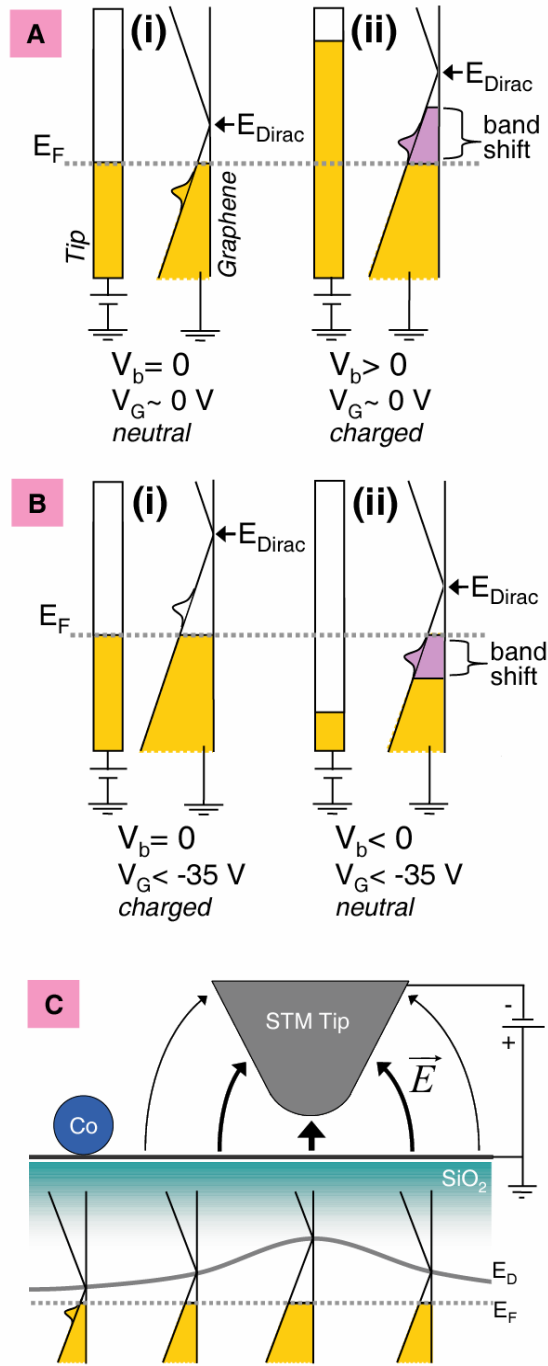